\documentclass[journal]{IEEEtran}
\usepackage{graphicx}
\usepackage{cite}
\usepackage{picinpar}
\usepackage{amsmath}
\usepackage{url}
\usepackage{flushend}
\usepackage[latin1]{inputenc}
\usepackage{colortbl}
\usepackage{soul}
\usepackage{multirow}
\usepackage{pifont}
\usepackage{color}
\usepackage{alltt}
\usepackage{hyperref}
\usepackage{enumerate}
\usepackage{epstopdf}
\usepackage{pbox}
\usepackage{enumitem}
\usepackage{fancyhdr}

\usepackage{bm}
\usepackage{amssymb}
\usepackage{stfloats}
\usepackage{algorithmic}
\usepackage[ruled,vlined]{algorithm2e}
\usepackage[table]{xcolor}

\usepackage{wasysym}
\usepackage{diagbox}
\usepackage{tabularx}
\usepackage{subfigure}
\usepackage{caption}
\usepackage{color}
\usepackage{makecell}
\usepackage{booktabs}
\usepackage{multirow}

\begin{document}

\title{Agent-OSI: An Interoperability Architecture for Communication and Settlement in the Decentralized Internet of Agents}

\author{Wenxin~Xu,~Taotao~Wang,~\IEEEmembership{Member,~IEEE},~Yihan~Xia,~Shengli~Zhang,~\IEEEmembership{Senior~Member,~IEEE},~and~Soung~Chang~Liew,~\IEEEmembership{Fellow,~IEEE} %
\thanks{W.~Xu, T.~Wang, Y.~Xia and S. Zhang are with the College of Electronics and Information Engineering, Shenzhen University, Shenzhen, China (emails: 2022280280@email.szu.edu.cn, ttwang@szu.edu.cn, xiayihan2023@email.szu.edu.cn, zsl@szu.edu.cn). \emph{(Corresponding author: Taotao Wang.)}}

\thanks{S. Liew is with the Department of Information Engineering, The Chinese University of Hong Kong, Hong Kong (email: soung@ie.cuhk.edu.hk). }

}

\maketitle

\begin{abstract}
Large Language Models (LLMs) are accelerating the shift from an Internet of information to an Internet of Agents (IoA), where autonomous entities discover services, negotiate, execute tasks, and exchange value. Yet today's agents are still confined to platform silos and proprietary interfaces, lacking a common stack for interoperability, trust, and pay-per-use settlement. This article proposes \textit{Agent-OSI}, a functional interoperability architecture for a decentralized IoA, whose core contribution is agent-to-agent (A2A) communication and a Web-compatible, backend-agnostic settlement protocol built on HTTP 402 (Payment Required); identity, verifiable execution, and semantic orchestration are treated as boundary layers with interfaces to existing standards. We treat HTTP 402 as an application-layer challenge--response primitive---analogous to HTTP 401 for authentication---whose settlement backend (escrow contract, payment channel, or signed off-chain receipt) is a pluggable choice, instantiated via a blockchain escrow in our prototype. We implement a prototype and evaluate its communication and settlement performance. Results show that, for generative workloads, end-to-end latency is dominated by task execution rather than settlement confirmation, and that keeping negotiation and delivery off the settlement backend reduces per-session settlement cost by approximately 51\% relative to a more on-chain baseline.

\end{abstract}

\begin{IEEEkeywords}
Internet of Agents, agent-to-agent communication, decentralized settlement, HTTP 402, verifiable execution
\end{IEEEkeywords}

\markboth{}{}

\section{Introduction}\label{s:intro}

\IEEEPARstart{T}{he} rapid advancement of Large Language Models (LLMs) is accelerating the transition from an Internet of information to an Internet of Agents (IoA), where autonomous entities discover services, negotiate terms, execute tasks, and exchange value with minimal human intervention. Recent IoA research has demonstrated heterogeneous agent collaboration \cite{chen2025internet}, studied capability discovery and risk-aware verification for agentic peer-to-peer networks \cite{wang2026agentic_p2p}, and surveyed enabling technologies and open challenges across communication, identity, security, and incentives \cite{wang2025internet}. As a result, agents increasingly behave as networked principals interacting across vendors and administrative domains.

Despite progress in multi-agent frameworks and emerging protocols, deployments remain platform-centric and tied to proprietary interfaces, leaving interoperability fragmented \cite{chen2025internet,wang2025internet}. Similar barriers appear in the broader Web~3.0 transition, where decentralization and user control are constrained by scalability, interoperability, usability, and trust challenges \cite{cao2025web}.

For an open and decentralized IoA, this creates a practical gap: agents may be capable in isolation, yet lack shared mechanisms to reliably (i) \textit{identify} each other and establish cross-domain trust \cite{wang2025security_ioa,mazzocca2025did_vc_survey}, (ii) \textit{communicate} with secure asynchronous semantics suitable for offline settings \cite{rfc9420_mls}, (iii) \textit{negotiate and pay} per request without a shared billing provider \cite{interledger_ilp,ietf_http_semantics}, and (iv) \textit{produce verifiable evidence} of execution and delivery for auditing and dispute hooks \cite{slsa_spec,intoto_spec}. Without shared protocols and explicit interfaces, cross-platform collaboration remains the exception.

\begin{figure*}[t]
    \centering
    \includegraphics[width=0.9\linewidth]{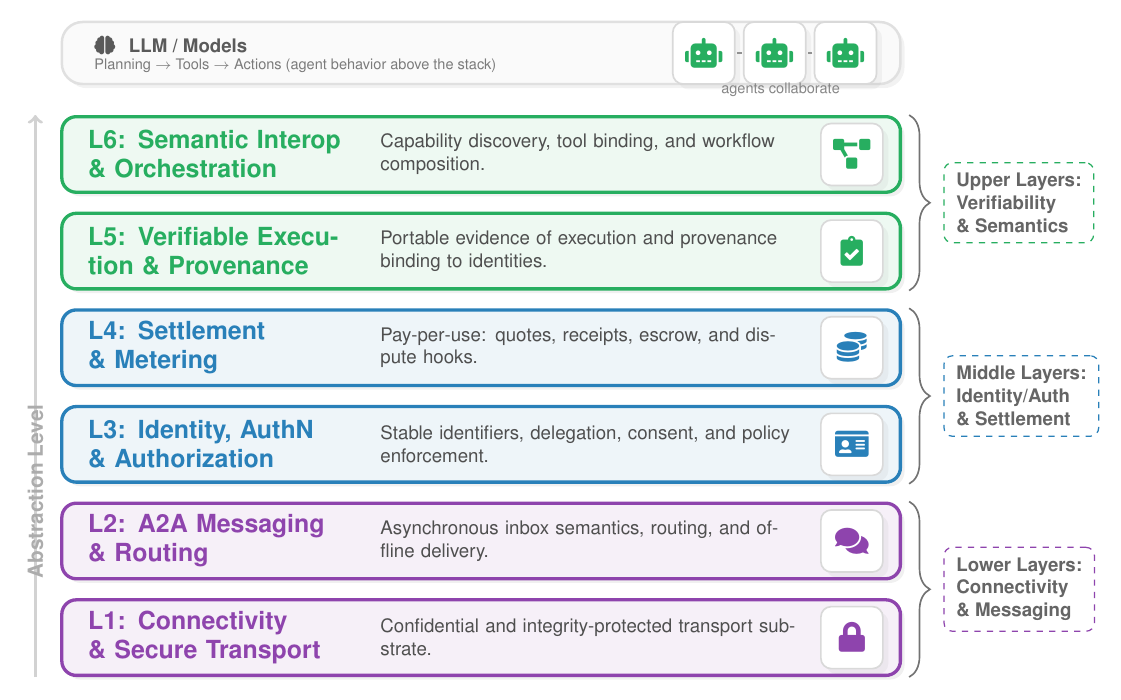}
    \caption{The Agent-OSI reference architecture, built on top of today's Internet stack. Representative protocols per layer (illustrative, non-exhaustive): L6 MCP/A2A/OpenAPI/JSON~Schema; L5 SCITT/in-toto/SLSA/C2PA; L4 HTTP~402/ILP/escrow contracts; L3 W3C~DID/VC/OAuth~2.0/OIDC; L2 MLS/DIDComm/libp2p/Nostr; L1 IP/TCP/TLS/QUIC.}
    \label{fig:agent_osi}
\end{figure*}

To address this gap, we draw inspiration from the classical Internet: interoperability at scale emerged from layering with clean interfaces and separable trust assumptions. We propose \textit{Agent-OSI} (illustrated in Fig.~\ref{fig:agent_osi}), a functional interoperability architecture above TCP/IP---not a new network-layer protocol---for a decentralized IoA; its core contribution is agent communication and settlement (L1/L2/L4), with identity (L3), verifiable execution (L5), and semantic orchestration (L6) as boundary layers over existing standards. Our goal is a deployable architecture with explicit cross-layer interfaces so heterogeneous agent frameworks can interoperate without a single platform as the trust anchor.

A central element of Agent-OSI is a Web-compatible settlement interface for pay-per-use services. We use HTTP 402 (Payment Required) as an application-layer challenge--response primitive (analogous to HTTP 401 for authentication) carrying a machine-verifiable quote and receipt-verification rules, rather than introducing a new network-layer protocol \cite{ietf_http_semantics}. This enables a client agent to pay per request without proprietary billing APIs, while supporting replay protection, auditability, and dispute hooks.

Specifically, this article makes two contributions:
\begin{itemize}
    \item  We propose \textit{Agent-OSI}. Our novelty is not the individual mechanisms Agent-OSI reuses (DID, OAuth, MLS, escrow contracts), but the systematization of HTTP 402 as a backend-agnostic \textit{Payment Challenge} primitive (L4) that (i) binds a machine-verifiable quote and receipt to a specific request via a nonce and request hash to prevent replay, (ii) triggers heterogeneous settlement backends (e.g., escrow, payment channels, L2 rollups, or signed off-chain receipts) through a single unified interface, and (iii) cryptographically binds payments to L5 execution provenance so that settlement cannot be equivocated from what was actually executed.

    \item We implement a working prototype, \textit{AgentMarket}, using commodity components (asynchronous messaging, ECDSA identities, pluggable settlement, signed provenance, and content-addressed delivery), and evaluate its communication and settlement performance. Results show that end-to-end latency for generative workloads is dominated by task execution rather than settlement confirmation, and that keeping negotiation and delivery off the settlement backend while keeping receipts verifiable reduces per-session on-chain settlement cost by approximately 51\% compared with a more on-chain baseline in our prototype setting.
\end{itemize}

\section{Why Decentralized Agent Networks?}
Before introducing the architecture, we clarify why the IoA benefits from a \emph{decentralized} networking model rather than relying solely on centralized marketplaces. Here, ``decentralized'' does not mean replacing the Internet; it means avoiding single-platform trust anchors for identity portability, messaging reachability, and cross-domain value exchange \cite{wang2025internet}. This yields four requirements. \emph{Interoperability}: most agent ecosystems couple discovery, tool invocation, and billing to provider-specific runtimes, which does not scale across administrative domains; an open IoA instead needs protocol-level machine-readable capability descriptions and durable asynchronous A2A messaging \cite{rfc9420_mls}. \emph{Sovereignty}: an agent must act as a principal---authenticating requests, holding assets, and delegating---with stable identifiers that survive provider changes rather than platform-issued ones \cite{mazzocca2025did_vc_survey}. \emph{Trust}: as agents initiate actions with economic consequences, the model shifts from ``trust the platform'' to ``verify the counterparty'' \cite{wang2025security_ioa}, making portable, machine-verifiable receipts and provenance preferable to opaque enforcement. \emph{Deployability}: the architecture must be incremental, running on today's Internet transports and reusing widely deployed security mechanisms, on a ``deploy-now, harden-later'' path where lightweight evidence is adopted early and stronger verification is introduced as the ecosystem matures.

\section{The Agent-OSI Architecture}\label{sec:agent_osi}

To standardize the development and interconnection of autonomous agents, we propose \emph{Agent-OSI}, a reference architecture for a decentralized IoA. Rather than redefining Internet protocols, Agent-OSI is designed to run \emph{on top of} today's TCP/IP-based network and to isolate agent-specific requirements into six functional layers. This separation supports incremental adoption and eases alignment with standardization efforts spanning networking, identity, and decentralized systems. As introduced in Section~\ref{s:intro}, the subsections below detail L1, L2, and L4 as the core contribution, while L3, L5, and L6 are boundary layers with interfaces to existing standards.

Fig.~\ref{fig:agent_osi} summarizes each layer and its core responsibility. Agent-OSI is inspired by OSI, but it is not a strictly encapsulated stack. While upper layers predominantly consume services from lower layers, some security- and economics-critical properties require explicit cross-layer bindings. In particular: (i) L4 settlement conditions may depend on L5 execution evidence and provenance, and (ii) L4 metering and pricing may depend on L6 capability declarations and semantic contracts. We make these dependencies explicit via signed, machine-verifiable artifacts (e.g., quote, receipt, and provenance objects) and treat them as controlled cross-layer interfaces rather than ad hoc coupling. These bindings prevent equivocation between what was executed (L5/L6) and what was paid for (L4). Concretely, what distinguishes a controlled interface from tight coupling is substitutability: any conformant implementation that produces an artifact of the expected type and schema---e.g., a different L5 evidence backend (Section~III-E) or a different L6 capability manifest---can be substituted without modifying the consumer's (L4's) verification logic, because that logic only inspects the signed artifact, not the internal mechanism that produced it. Agent-OSI is therefore best understood as a functional interoperability architecture, in which L4, L5, and L6 interact only through these artifact contracts rather than through implicit shared state.

Agent-OSI is intended to complement ongoing industry and standards efforts---not to replace Internet protocols. At the communication layer, it can reuse secure messaging substrates such as MLS~\cite{rfc9420_mls}. For identity and authorization, Agent-OSI aligns with decentralized identity primitives~\cite{mazzocca2025did_vc_survey} while remaining compatible with widely deployed Web authorization patterns such as OAuth~2.0~\cite{rfc6749_oauth2}. For settlement, we view HTTP~402 as a Web-compatible payment challenge~\cite{ietf_http_semantics} that can trigger interoperable payment rails such as ILP~\cite{interledger_ilp} or escrow-style receipts verifiable against on-chain events~\cite{evm_yellowpaper}. Finally, verifiable execution and provenance are designed to interoperate with established supply-chain and provenance practices (e.g., in-toto and SLSA)~\cite{intoto_spec,slsa_spec}.

\subsection{L1: Connectivity and Secure Transport}
L1 provides baseline connectivity and secure transport. Agent-OSI does not replace the Internet stack; instead, it assumes standard IP routing and widely deployed transport and security mechanisms (e.g., QUIC/TLS) as the substrate. This choice keeps the proposal deployable and allows higher layers to focus on agent-specific semantics rather than packet delivery. L1 deliberately bundles classical network, transport, and presentation-layer functionality (IP, TCP, TLS/QUIC) into a single connectivity dependency: Agent-OSI uses ``layer'' in a functional sense and does not re-partition the classical Internet stack, but treats it as a substrate.

\subsection{L2: A2A Messaging and Routing}
L2 provides A2A\footnote{In this article, ``A2A'' is used as a generic abbreviation for agent-to-agent messaging (the communication function) and does not refer to any particular protocol. ``Google A2A'' instead denotes Google's specific proposal that defines application-level message formats and interoperability conventions for agent-to-agent interactions, which is an L6 protocol in our architecture.} messaging with asynchronous delivery semantics. Compared to conventional synchronous REST calls, agents often require durable inboxes, offline delivery, and conversation threading. This layer defines message envelopes (sender/receiver identifiers, timestamps, nonces, thread IDs, and signatures) and uses an overlay messaging network (e.g., XMTP or libp2p) to route messages without assuming continuous availability of either party.

Durable inboxes raise a storage-cost question: who pays to retain a message until the recipient agent comes online? We identify three non-exclusive models compatible with Agent-OSI's L2 interface: \emph{sender-prepaid storage}, where the sending agent attaches a small, metered storage fee (itself expressible as an L4 payment challenge) to cover retention until delivery or expiry; \emph{recipient-hosted inboxes}, where the receiving agent or its operator hosts and pays for its own inbox, at the cost of reduced availability if the recipient is offline for extended periods; and \emph{overlay-subsidized storage}, where a messaging overlay amortizes storage cost across many users in exchange for a flat access fee. Our prototype uses recipient-hosted inboxes; selecting among these models under adversarial and economic constraints is future work.

\subsection{L3: Identity, Authentication, and Authorization}
L3 defines how agents are identified and authorized to act, reusing decentralized identifiers (DIDs) and Web authorization mechanisms rather than defining new ones. Identifiers must be stable across platforms and support key rotation, recovery, and revocation; the layer also covers authentication (signed requests) and authorization (policy and consent). On-chain accounts such as ERC-6551 are an optional asset-holding container, not a complete identity system.

\subsection{L4: Settlement and Metering}
L4 supports value exchange for pay-per-use agent services. It covers pricing, usage metering, payment challenges, receipts, and escrow-based settlement flows.
Metering semantics are defined by the service's capability/contract (L6) and cryptographically bound to the quote/receipt. We treat HTTP 402 (Payment Required) as an application-level payment challenge (similar in spirit to HTTP 401 for authentication): a service may return 402 to indicate that a request is valid but requires payment before execution \cite{ietf_http_semantics}.

In Agent-OSI, a 402 response includes a structured quote and a receipt specification so that a client agent can pay and present proof of payment in a standard way. The intent is not a new network-layer protocol, but a thin, Web-compatible interface that can trigger multiple settlement backends while remaining interoperable across agent implementations.

\subsubsection{402 Payment Challenge Message}\label{sssec:402_challenge}
A 402 response carries a machine-readable payment challenge with the minimum fields needed for interoperability: price and currency (e.g., \texttt{0.25 USDC}); payee identity (DID and/or on-chain address) with a signature binding the quote to the payee; payment network and method (chain ID, token contract, escrow address); an expiration time and nonce to prevent replay; and the receipt format and verification rules (on-chain event proof, signed receipt, or L2 receipt).

\subsubsection{Receipt and Verification}\label{sssec:receipt_verification}
After payment, the client provides a receipt that the service can verify prior to execution or prior to releasing escrow. Depending on the deployment, a receipt can be: (i) an on-chain event inclusion proof (transaction hash plus event log), (ii) a signed receipt issued by a payment processor, or (iii) an L2 rollup receipt. Agent-OSI only requires that the receipt be verifiable and bound to the original quote (nonce and request hash), so that payment proofs cannot be reused across different requests. The receipt can additionally be bound to an L5 provenance object (or its hash) to support escrow release conditions without placing delivery on-chain.

\subsubsection{Integrating Payment Channels and L2 Rollups}
Because the receipt-type field (Section~\ref{sssec:402_challenge}) is backend-agnostic, integrating a payment channel or an L2 rollup requires only a new receipt-verification adapter on the service side, not a change to the client-facing protocol. For an L2 rollup, the escrow event proof is replaced by a rollup inclusion proof (setting the receipt type to \texttt{l2\_rollup}); for a payment channel, by a signed channel-balance update checked against the channel's latest on-chain checkpoint. In both cases the quote structure, the nonce/\texttt{requestHash} binding, and the escrow-release condition remain unchanged.

\subsubsection{Adoption Barriers}
Practical adoption of 402-based settlement faces at least three barriers. First, \emph{wallet interoperability}: clients must hold keys compatible with the declared payment network/method, and cross-wallet or cross-chain payment UX remains immature relative to existing Web payment flows. Second, \emph{dispute handling}: a 402-triggered escrow needs an explicit, minimal dispute interface (timeouts, admissible evidence such as L5 provenance, and an arbitration role) to resolve non-delivery or quality disagreements. Third, \emph{integration with existing Web/API infrastructure}: reverse proxies, API gateways, and billing middleware would need to recognize and forward 402 challenges and receipts, which is a deployment/tooling gap rather than a protocol limitation.

\subsubsection{Attack Surface}
Introducing payment and metering expands the adversarial surface in at least three ways, each constrained by explicit bindings in our design. \emph{Forged receipts} are rejected unless they decode to an on-chain (or backend-specific) event matching the quote's payer/payee, amount, and expiry. \emph{Quote manipulation} is prevented because every quote is signed by the payee and bound to a nonce and a \texttt{requestHash}, so a modified or replayed quote fails verification against the original request. \emph{Incorrect or dishonest metering} is constrained because metering semantics are declared in the L6 capability manifest and cryptographically referenced by the quote, so a service cannot silently change its metering basis after a quote has been issued; detecting a service that meters honestly according to its declared policy but misrepresents actual usage remains an open problem (Section~VI).

\begin{figure*}[htbp]
    \centering
    \includegraphics[width=0.8\textwidth]{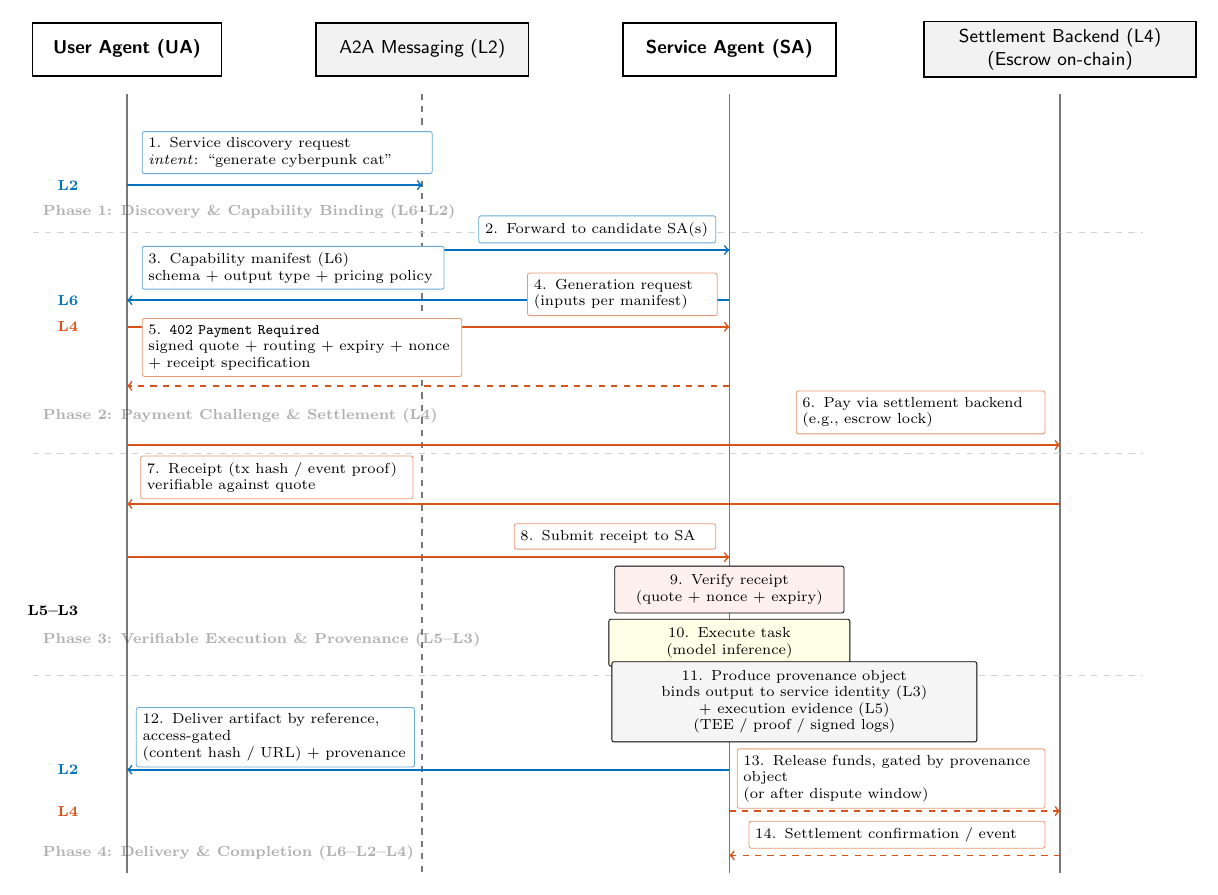} 
    \caption{Sequence diagram of the cross-layer interaction. The workflow coordinates A2A messaging, payment challenge and settlement, verifiable execution evidence, and delivery.}
    \label{fig:sequence_flow}
\end{figure*}

\subsection{L5: Verifiable Execution and Provenance}
L5 provides evidence that a service was executed under agreed conditions and that outputs are bound to a verifiable provenance record. Depending on the threat model and performance constraints, this may rely on TEE attestations, zero-knowledge proofs for specific computations \cite{wang2026zkfl_commsmag}, or signed execution logs that enable auditing. L5 is designed to produce portable proof artifacts that can be consumed by L4 for settlement and by L6 for downstream reuse and composition.

In practice, L5 admits multiple instantiations with different assurance levels (e.g., TEE attestations or zero-knowledge proofs), each imposing distinct trust assumptions and performance costs. Concretely, these backends trade off overhead against assurance quite differently: SGX-style TEE attestation typically adds low, near-constant per-call latency, but requires trusting the enclave's hardware root of trust and remains susceptible to known side-channel classes; ZKML proof generation for neural-network inference currently incurs orders-of-magnitude higher latency and memory cost than the underlying computation, making it more suitable for small or infrequently verified models than for large generative workloads such as our SDXL case study (Section~IV); and signed execution logs, used in our prototype, add negligible overhead but provide auditability rather than cryptographic non-repudiation of correctness. Agent-OSI does not mandate a single mechanism; instead, it standardizes the interface for producing and consuming provenance artifacts so that settlement (L4) and orchestration (L6) can be configured according to the required trust model.

\subsection{L6: Semantic Interoperability and Orchestration}
L6 is the top layer responsible for interoperability and composition. It standardizes capability descriptions and tool interfaces so that agents can discover, bind, and invoke services without bespoke integrations. It also defines structured context objects and provenance-aware knowledge exchange, enabling multi-agent workflows and orchestration frameworks to coordinate tasks across heterogeneous providers. Protocols such as MCP and schema-based API descriptions fit naturally here, alongside workflow engines for multi-agent coordination. We note that L6, as specified here, standardizes syntactic interoperability---machine-checkable schemas for capability manifests and tool interfaces---rather than deeper semantic alignment across agents. Achieving such alignment, potentially through emergent or learned semantic protocols between agents, is an open research direction that Agent-OSI's L6 interface is designed to accommodate but does not itself solve.

\section{Case Study: Autonomous Content Creation}
\label{sec:case_study}

To illustrate how Agent-OSI operates end-to-end, we use an autonomous content creation session as a running example, also the concrete interaction implemented and evaluated in our prototype \textit{AgentMarket}: a \textit{User Agent (UA)} commissions a \textit{Service Agent (SA)} to generate an image, following the discovery$\rightarrow$payment$\rightarrow$execution$\rightarrow$delivery workflow of Fig.~\ref{fig:sequence_flow} in four phases.

\emph{Phase 1 --- Discovery and capability binding (L6--L2).} The UA forms an intent (``generate a cyberpunk cat image'') and discovers candidate services over the A2A messaging layer (L2); the SA replies with a capability manifest (L6) specifying input schema, output type, and pricing, letting the UA bind without hard-coded integration. The UA discovers candidate SAs through a lightweight registry that indexes signed capability manifests, rather than a hard-coded address. Because manifests are signed under L3 identities, the registry is an index rather than a trust root---a misbehaving registry can omit but not forge entries---and the same interface admits decentralized (gossip-based) discovery, at the cost of weaker scalability and greater participation-metadata exposure (a linkability concern we discuss in Section~VI).

\emph{Phase 2 --- Payment challenge and settlement (L4).} The SA responds to the generation request with \texttt{402 Payment Required} (signed quote, routing, expiry, nonce, receipt spec); the UA pays through the configured backend and returns a receipt verifiable against the quote.

\emph{Phase 3 --- Verifiable execution and provenance (L5--L3).} After verifying the receipt, the SA executes the task and produces a provenance artifact binding the output to the service identity (L3) and execution evidence (L5); the evidence form (TEE attestation, ZK proof, or signed logs) is workload- and threat-model-dependent.

\emph{Phase 4 --- Delivery and completion (L6--L2--L4).} The artifact is delivered over L2 by reference (content hash/URL) plus the provenance object; if escrow is used, the contract releases funds per a completion rule referencing L5 evidence and L6 contract terms, keeping delivery largely off-chain. Fair exchange is preserved by what is disclosed at each step, not by their order: delivery ``by reference'' gives the UA only a content hash/CID bound to the provenance object, while the underlying content is access-gated and released only once the escrow-release condition---evaluated against that same object---is satisfied. Receiving the reference thus does not let the UA obtain usable content ahead of the SA's payout.

\section{Prototype Implementation and Evaluation}
\label{s:performance_evaluation}

We built a proof-of-concept prototype system, \textit{AgentMarket}, to validate Agent-OSI for pay-per-request agent services and investigate it experimentally. The code of the prototype is publicly available at \url{https://github.com/plan-lab-szu/agent-market}.

\subsection{Prototype Implementation}
The AgentMarket prototype implements an end-to-end session covering capability advertisement and binding, a 402-style payment challenge with escrow settlement, provenance generation, and content-addressed delivery (following Section~\ref{sec:case_study} and Fig.~\ref{fig:sequence_flow}).

\subsubsection{Stack Instantiation}
AgentMarket follows the UA--SA interaction in Section~\ref{sec:case_study} using two lightweight Python services. The UA discovers the SA via a machine-readable capability manifest; the SA returns signed quotes/payment challenges, verifies escrow receipts, executes the task, and returns delivery plus provenance.

Each layer is instantiated with commodity components: a JSON-Schema capability manifest (L6), ECDSA keypairs as DID-compatible identities (L3), XMTP for asynchronous inbox messaging (L2), an EVM escrow contract for settlement and receipt verification (L4)~\cite{evm_yellowpaper} deployed on a local Anvil testbed (fixed 2\,s block time), and a portable provenance artifact derived from a canonical execution-log hash signed by the SA (L5).

\subsubsection{Scope and Reproducibility Notes}
The prototype validates workflow correctness and interface interoperability rather than optimizing a specific verifiable-execution backend; accordingly, L5 uses signed, auditable provenance logs (extendable to TEE attestations or ZK proofs). For reproducibility, we fix the manifest schema and quote function, canonicalize JSON before hashing, and log all session artifacts (quotes, receipts, provenance, and message transcripts). Delivery uses standard IPFS pinning: once uploaded, an artifact is publicly retrievable by anyone holding its CID. The access-gating described in Section~IV, Phase~4 (e.g., client-side encryption with a key released only upon escrow confirmation, or a permissioned delivery service) is an architectural option that the current prototype does not enforce; closing this gap is left for future work.

\subsubsection{Messages, Receipts, and Provenance Objects}
AgentMarket exchanges a small set of signed objects over L2: message envelopes for asynchronous A2A delivery, signed quotes carried by 402 challenges, escrow receipts verifiable from on-chain events, and an L5 provenance artifact bound to the delivered content.

Replay resistance relies on two session-wide bindings: \texttt{requestHash} (SHA-256 over canonicalized request JSON) and \texttt{quoteId} (derived from the signed quote and \texttt{requestHash}). A payment is accepted only if the decoded escrow event matches these bindings (as well as payer/payee, amount, and expiry). The provenance artifact records \texttt{requestHash}/\texttt{quoteId}, references the verified receipt (e.g., transaction hash), and binds the output via its CID together with a signed execution-log hash.

\subsubsection{Model and Hardware Configuration}
For a representative heavy-compute workload, the SA runs Stable Diffusion XL (SDXL~1.0) \cite{podell2023sdxl} via HuggingFace Diffusers. Unless noted, we fix $1024\times1024$ resolution, 30 denoising steps, and a fixed random seed. Experiments run on a single-GPU machine (RTX4090 24GB, CUDA 13.0). We include packaging and IPFS upload/pinning in the execution-time breakdown.

\begin{figure*}[htbp]
    \centering
    \includegraphics[width=1\textwidth]{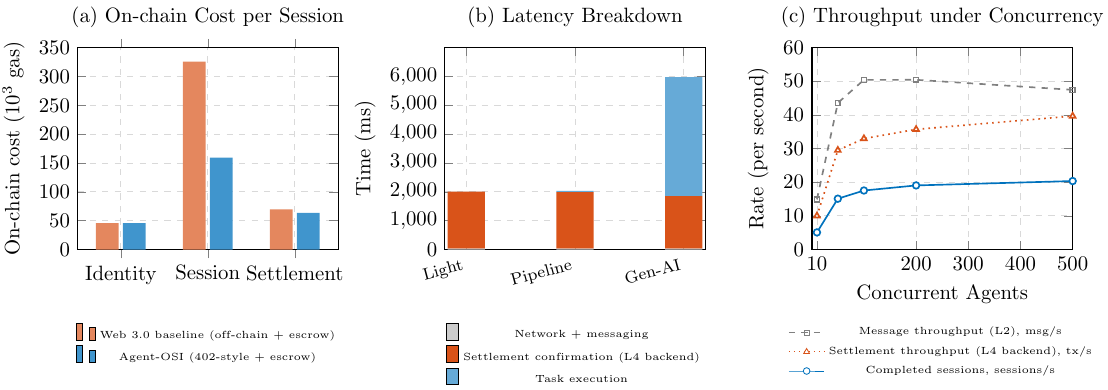}
    \caption{Performance of the Agent-OSI prototype. (a) \textit{Cost:} on-chain EVM execution cost per completed paid session (sum of \texttt{gasUsed}; network fees excluded); Web 2.0 baseline (off-chain billing) has zero on-chain cost and is omitted. (b) \textit{Latency:} end-to-end breakdown (messaging/network, confirmation, execution/delivery) for Light (no-gen), Pipeline ($K$-step, IPFS/CID), and GenAI (image/LLM, IPFS/CID). (c) \textit{Throughput:} steady-state message (L2), settlement (L4 confirmed escrow tx), and completed-session throughput under increasing concurrency (UAs$\rightarrow$1 SA).}
    \label{fig:evaluation}
\end{figure*}

\subsection{Experimental Evaluation}
Fig.~\ref{fig:evaluation} summarizes cost, latency, and throughput. Unless otherwise stated, we run 100 sequential trials and report medians with p10--p90 ranges. We attribute bottlenecks by separating on-chain settlement cost/confirmation from off-chain delays (messaging, receipt verification, execution, and IPFS upload).

\subsubsection{Cost Evaluation (Fig.~\ref{fig:evaluation}a)}
We measure on-chain cost as the gas used over the minimal set of EVM transactions per paid session (network fees excluded), comparing Agent-OSI against a Web 3.0 dAPP escrow baseline that anchors session metadata on-chain (a centralized Web 2.0 baseline has zero on-chain cost and is omitted). Identity registration costs are constant ($\approx$46k gas) across both; by keeping service gating and negotiation off-chain and relying on receipt-bound verification, Agent-OSI reduces session-related gas by $\approx$51\% ($\approx$326k$\to\approx$159k gas).

\subsubsection{Latency Evaluation and Boundary Conditions (Fig.~\ref{fig:evaluation}b)}
We evaluate three workloads: \textit{Light (no-gen)} returns a small inline JSON result; \textit{Pipeline ($K$-step)} produces $K{=}5$ fixed-size artifacts uploaded to IPFS; and \textit{GenAI} performs a generative task (SDXL). Fig.~\ref{fig:evaluation}(b) decomposes latency into messaging, settlement confirmation (on a fixed 2\,s block-time Anvil testbed), and execution. Messaging is consistently low ($\approx$30--35 ms). For \textit{GenAI}, task execution ($\approx$4122 ms) dominates, so settlement confirmation ($\approx$1820 ms, $\approx$30\% of total) is masked; for \textit{Light}, settlement confirmation is the bottleneck ($\approx$98\%), indicating that simple high-frequency tool use needs asynchronous or batch settlement.

\subsubsection{Throughput and Metric Definitions (Fig.~\ref{fig:evaluation}c)}
We report message throughput (L2), settlement throughput (L4), and completed-session throughput. We run fixed-duration load tests with increasing concurrency (10 to 500 agents). As shown in Fig.~\ref{fig:evaluation}(c), message throughput saturates first, peaking at $\approx$50 msg/s. Settlement throughput scales with concurrency but is bounded by RPC and chain limits, reaching $\approx$40 tx/s at 500 concurrent agents. Consequently, the completed-session throughput reaches a maximum of $\approx$20 sessions/s. This aligns closely with the theoretical bound where $\text{sessions/s} \approx \frac{\text{tx/s}}{2}$ (since each session requires lock and release transactions), confirming that the system's scalability is currently constrained by the underlying blockchain settlement rate rather than the agent messaging layer. This ceiling is not fundamental to the 402 interface, however: it reflects a single chain settling one lock/release pair per session, and the same interface can raise it through batched or asynchronous settlement, payment-channel or L2/rollup backends (Section~III-D3), or cross-chain sharding. The prototype is thus well suited to moderate-value, moderate-frequency sessions such as our SDXL case study, whereas high-frequency, low-value tool calls would require one of these optimizations, which we leave to follow-up work.

\textit{Scope of these measurements.} These results use a single local Anvil chain with co-located agents---favorable conditions that isolate the settlement bottleneck. A wide-area, public-chain deployment would generally show \emph{higher} latency and \emph{lower} throughput; the gas-cost savings are fixed by the contract logic and carry over, but the latency/throughput figures characterize the architecture's overhead structure, not absolute production performance. Wide-area, multi-chain, and heterogeneous-trust-domain evaluation is future work.

\section{Challenges and Open Issues}
\label{sec:challenges}

Agent-OSI is a reference architecture rather than a monolithic protocol. While it clarifies responsibilities across layers (Fig.~\ref{fig:agent_osi}), deploying a decentralized Internet of Agents at scale still faces several open issues where \emph{cross-layer interfaces} matter as much as individual mechanisms.

\emph{1) Privacy and linkability across messaging, settlement, and evidence (L2/L4/L5)}.
Even with encrypted payloads (L1), real deployments leak metadata. L2 reveals timing and conversation structure; L4 receipts and payment challenges may be publicly observable; and L5 provenance can unintentionally expose inputs, tools, or workflow structure. Key directions include: (i) privacy-preserving asynchronous A2A overlays (e.g., reducing traffic analysis while keeping inbox semantics), (ii) receipts with selective disclosure that still support replay protection and auditing, and (iii) provenance minimization so verifiers can check completion or policy compliance without learning unnecessary details.

We note a direct tension between the cross-layer bindings introduced in Section~III-D for replay protection (nonce/\texttt{requestHash} linking quote, receipt, and provenance) and used for fair exchange in Section~\ref{sec:case_study}, Phase~4, and the privacy/linkability goals discussed here: the same binding that lets a verifier confirm that a payment corresponds to an execution also lets an observer link payer, service, and time window across L2, L4, and L5 artifacts. Selective-disclosure receipts and provenance minimization would need to preserve the verifiability of these bindings---e.g., via zero-knowledge proofs of correct binding---rather than removing them outright, since removing them would reopen those fair-exchange and replay concerns. Reconciling these two requirements is, to our knowledge, an open problem.

\emph{2) Identity, delegation, and key management at agent scale (L3)}.
Open IoA requires identities that survive provider changes and support rotation/recovery, plus explicit delegation (agents acting for users/orgs) and least-privilege authorization. In practice, deployments will likely mix decentralized identifiers with OAuth/OIDC-style usability patterns; defining safe interoperability profiles and clear trust anchors for such compositions remains an open standardization and engineering problem.

\emph{3) Settlement semantics and dispute hooks for 402-style flows (L4 with bindings to L5/L6)}.
Treating HTTP 402 as a payment challenge yields a Web-compatible interface, but semantics must be crisp. Metering needs precise definitions (per request/tokens/steps/resources), and receipts must bind to a specific quote and request (nonce + request hash) to support idempotent retries and prevent replay. Disputes are especially under-specified: quality claims, non-delivery, and partial execution require a \emph{minimal dispute interface} (timeouts, admissible evidence, arbitration roles) that works across settlement backends while keeping delivery off-chain.

\emph{4) What does ``verifiable execution'' mean in practice? (L5 with consumable statements)}.
TEE attestations, ZK proofs, and signed logs provide different assurance levels, yet are often over-interpreted. A practical path forward is to standardize \emph{evidence statements and schemas} (what is asserted, under what assumptions), enable composable provenance across multi-agent workflows, and adopt scalable verification strategies (e.g., layered assurance and sampling audits) so high-assurance mechanisms are reserved for high-value sessions.

Overall, these challenges suggest that the next step for an open IoA is not a single ``best'' protocol, but \emph{interoperability profiles} that specify required artifacts (quote/receipt/provenance), minimum security properties, and conformance tests across layers.

\section{Conclusion}

The shift from chatbots to autonomous agents will reshape how digital services are discovered, composed, and paid for. Without shared communication and settlement interfaces, however, the Internet of Agents risks fragmenting into closed ecosystems of proprietary APIs. We introduced Agent-OSI, a functional interoperability architecture centered on agent communication and settlement (L1/L2/L4), with identity, verifiable execution, and orchestration as boundary layers. We further showed how HTTP 402 can serve as an application-level payment challenge for pay-per-use workflows with verifiable receipts, while keeping negotiation and delivery largely off-chain. In our prototype, settlement confirmation is often masked by task execution time for generative workloads, and keeping the session off-chain is economically competitive with centralized marketplaces. Our results suggest that a backend-agnostic payment-challenge interface, cryptographically bound to execution provenance, is a practical and measurable step toward interoperable agent communication and settlement. We do not claim that Agent-OSI itself is already a settled standard for the Internet of Agents.

\bibliographystyle{IEEEtran}

\bibliography{ref.bib}

\end{document}